\documentclass[twocolumn,showpacs,preprintnumbers,amsmath,amssymb]{revtex4}
                                                                                
\usepackage{graphicx}% Include figure files
\usepackage{dcolumn}% Align table columns on decimal point
\usepackage{subfigure}
\usepackage{bm}% bold math
\def\be{\begin{equation}}
\def\ee{\end{equation}}
\def\bc{\begin{center}}
\def\ec{\end{center}}
                                                                                
\begin{document}
                                                                                
\title{Percolation in a kinetic opinion exchange model}% Force line breaks with \\

\author{Anjan Kumar Chandra}
\email{anjanphys@gmail.com}
\affiliation{%
Theoretical Condensed Matter Physics Division, Saha Institute of Nuclear
Physics, \\ 1/AF Bidhannagar, Kolkata-700064, India.}

%\date{\today}
                                                                                
\begin{abstract}
We study the percolation transition of the geometrical clusters in the square
lattice LCCC model (a kinetic opinion exchange model introduced by Lallouache {\em et al.}
in Phys. Rev. E {\bf 82} 056112 (2010)) 
with the change in conviction and influencing parameter.
The cluster comprises of the adjacent sites having an opinion value greater than or
equal to a prefixed threshold value of opinion ($\Omega$).
The transition point is different from that obtained for the transition of the
order parameter (average opinion value) found by Lallouache {\em et al.} Although the 
transition point varies with the change in the
threshold value of the opinion, the critical exponents for the percolation transition 
obtained from the data collapses 
of the maximum cluster size, cluster size distribution and Binder cumulant remain same. The exponents are also 
independent of the values of conviction and influencing parameters indicating the 
robustness of this transition. The exponents do 
not match with that of any other known percolation exponents (e.g. the static Ising,
dynamic Ising, standard percolation) and thus characterizes the LCCC model to belong to
a separate universality class.
\end{abstract}

                                                                                
\def\be{\begin{equation}}
\def\ee{\end{equation}}
\maketitle

\section{Introduction} 

Geometrical percolation transition has been a long studied subject 
\cite{Stauffer,Grimmett}. It is characterised by a set of universal critical
exponents, which describe the fractal properties of the percolating medium 
at large scales and sufficiently close to the transition. The exponents
only depend on the type of percolation model and on the spatial dimension. 
The occupancy of the sites or bonds of a percolating system
is controlled by a parameter and at a critical value of that parameter the 
cluster sizes (defined by the number of adjacent sites posessing a pre-defined common
feature) goes to infinity which we call percolation transition. This phenomenon
has been extensively studied for the thermal excitation of the two dimensional Ising
model and in this case the system undergoes percolation transition 
\cite{Muller,Stoll,Muller2,Binder} at the same critical temperature as the 
magnetization \cite{Coniglio1,Coniglio2}. In case of Ising model, the geometrical
cluster is defined by the adjacent sites consisting of parallel spins. The transition 
point differs in case of higher dimensions \cite{Muller,Heermann}. The percolation
exponents of the geometrical clusters are identical for the models belonging to
the same universality class (Ising and Z(3) symmetric models \cite{Santo1,Santo2}).
Recently the dynamical percolation transition has been studied for 2d Ising model
by applying pulsed magnetic field \cite{Biswas}. The critical exponents were different
from that of the static percolation transition associated with the thermal transition
of the Ising model indicating a different universality class. The distinct crossing 
point of the Binder cumulant of the order parameter for different system sizes at the 
transition point also characterizes the being of the dynamical pecolation transition 
in a different universality class. 

Study of social dynamics has been very popular in 
recent times and to capture the basic idea of consensus formation concepts of
statistical physics has been applied largely \cite{Castellano,Wiley}. A large number 
of 
models has been studied (voter model \cite{Holley,Liggett}, 
Sznajd model \cite{Sznajd} etc) so far.
In some models opinions have been considered as a continuous variable 
\cite{Hegselmann,Deffuant,Fortunato1,BCS,Lallouache1,Anirban}. The spreading of an opinion through the 
society may be compared with the percolation problem in physics and has been
studied for nonconsensus opinion model earlier \cite{Shao} and was found to belong
to the same universality class as the invasion percolation. 
In this paper we have studied the percolation 
transition of geometrical clusters in a recently proposed opinion model called
the LCCC model \cite{Lallouache1,Lallouache2} in which individuals exchange opinions
controlled by an influencing parameter and a conviction parameter, the values of which
are equal. The opinion of an individual is taken uniformly between $-1$ and $+1$
which changes by binary interactions, where an individual stays with his
own opinion upto a certain fraction $\lambda$ and takes a random fraction of
a part of another agent's opinion determined by the same parameter
(detailed discussion has been given in the next section). By Monte Carlo simulation 
it was found that below a critical value ($\lambda_c \approx 2/3$) of the conviction 
parameter the average 
opinion value remains zero, whereas above the critical value the average 
opinion value becomes non-zero. Some critical exponents characterising the 
transition in LCCC model and
some variants of the LCCC model were studied numerically \cite{Biswas2}.
A generalised version of this model was introduced by Sen \cite{Sen} in which
the influencing parameter and the conviction parameter were different.
A discrete version of the LCCC model has also been studied \cite{Biswas3}.

In this paper we have investigated the percolation transition of the geometrical 
clusters of the LCCC model assuming individuals are located
on the sites of a square lattice. We have defined $clusters$ as a group of adjacent 
sites
with opinion value equal to or above a preassigned threshold value ($\Omega$). 
The cluster sizes are controlled by the influencing  parameter $\lambda$ and for a 
fixed $\Omega$,
at a critical value of the influencing parameter $\lambda_c^p$, the percolation 
transition occurs. We determine the critical exponents by finite size scaling
analysis of the maximum cluster size. The value of the critical point 
decreases with decrease of $\Omega$ and coincides with that for the transition
point of $\lambda_c = 2/3$ (at which the average opinion diverges) as 
$\Omega \rightarrow 0.0$. But the critical exponents remain unaltered with
change in $\Omega$ and also differ with the exponents known for the
previously known models indicating the square lattice LCCC to belong to a separate 
university class. We have also investigated this percolation transition in case
of generalised LCCC model \cite{Sen}, where the conviction parameter ($\lambda$) is 
different 
from the influencing parameter ($\mu$) and once again found that although the 
critical point shifts depending on the values of $\Omega$, $\lambda$ and $\mu$, 
the critical exponents remain same.

The paper has been organised in the following manner : In Sec II we give a brief
description of both the LCCC model and the generalised LCCC model. In Sec III
we present the description of clusters and measure the critical exponents for
the square lattice LCCC model. In Section IV we measure the exponents for
the generalised LCCC model and finally in Sec V we have some discussions
regarding this transition and some conclusions.

%%%%%%%%%%%%%%%%%%%%%%%%%%%%%%%%%%%%%%%%%%%%%%%%%%%%%%%%%%%%%%%%%%%%%%%
\section{A brief description of the LCCC and generalised LCCC model}
%%%%%%%%%%%%%%%%%%%%%%%%%%%%%%%%%%%%%%%%%%%%%%%%%%%%%%%%%%%%%%%%%%%%%%%

The origin of this model is a multi-agent statistical model of closed economy
\cite{Chakrabarti} where $N$ agents exchange a fixed 
wealth through pair-wise interaction controlled by a ``saving" parameter.
Lallouache et. al. \cite{Lallouache1,Lallouache2} proposed a similar multiagent model
to describe the dynamics of opinion formation. The basic difference in this model is
that there is no constraint regarding the conservation of opinion. Let there are 
$N$ agents and each agent $i$ 
begins with an individual opinion $o_i \in [-1,+1]$. They exchange opinions between 
each other by binary interactions as follows :
\begin{eqnarray}
\label{eq:lccc}
 o_i(t+1) &=& \lambda[o_i(t) + \epsilon o_j(t)] \nonumber \\
 o_j(t+1) &=& \lambda[o_j(t) + \epsilon^\prime o_i(t)]
\end{eqnarray}
where $\epsilon$, $\epsilon^\prime$ are drawn randomly from uniform 
distributions in $[0,1]$. In this model the opinions are bounded i.e., 
$-1\le o_i \le +1$ for all $i$. Here the parameter $\lambda$ is interpreted as
``conviction" i.e., the power to retain someones own opinion. The second term 
signifies
the extent to which somebody get influenced by another. Here both the
conviction parameter and the influencing parameters are same 
and moreover they are identical for every individual. 
The opinion
exchange for the generalised LCCC model was as follows \cite{Sen}:
\begin{eqnarray}
\label{eq:glccc}
 o_i(t+1) &=& \lambda o_i(t) + \epsilon \mu o_j(t) \nonumber \\
 o_j(t+1) &=& \lambda o_j(t) + \epsilon^\prime \mu o_i(t)
\end{eqnarray}
where $\lambda$ is the conviction parameter and $\mu$ is the influencing parameter
with $o_i \in [-1,+1]$. 
The special case of $\lambda = \mu$ is the LCCC model.
The order parameter is the average opinion $O = |\sum_i o_i|/N$. Numerical 
simulations show that
the system stabilises into two possible phases : for any $\lambda \le \lambda_c$, 
$o_i=0$ $\forall i$, while for $\lambda > \lambda_c$, $O>0$ and $O \to 1$ as 
$\lambda \to 1$. In LCCC model $\lambda_c \approx 2/3$ is the critical point.
In case of the generalised model $\lambda_c$ depends on the value of $\mu$
and the mean field phase boundary is given by $\lambda = 1 -\mu/2$.
If we study these models on a square lattice, then also the critical points do not 
change. Some critical exponents characterising the transition in LCCC model and
some variants of the LCCC model were also studied numerically \cite{Biswas2}.

%%%%%%%%%%%%%%%%%%%%%%%%%%%%%%%%%%%%%%%%%%%%%%%%%%%%%%%%%%%%%%%%%%%%%%%
\section{Percolation on square lattice LCCC model}
%%%%%%%%%%%%%%%%%%%%%%%%%%%%%%%%%%%%%%%%%%%%%%%%%%%%%%%%%%%%%%%%%%%%%%%

\begin{figure}

\noindent \includegraphics[clip,width= 6cm, angle=270]{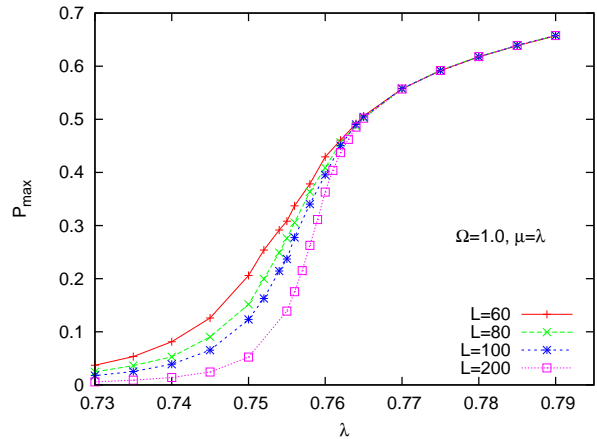}
\caption{(Color online) Maximum cluster size as a function of the conviction parameter  for four different system sizes ($L = 60, 80, 100$ and $200$) for the LCCC model i.e. $\mu = \lambda$ and threshold opinion value $\Omega = 1.0$.}
\label{fig:fg1}

\end{figure}

\begin{figure}

\noindent \includegraphics[clip,width= 6cm, angle=270]{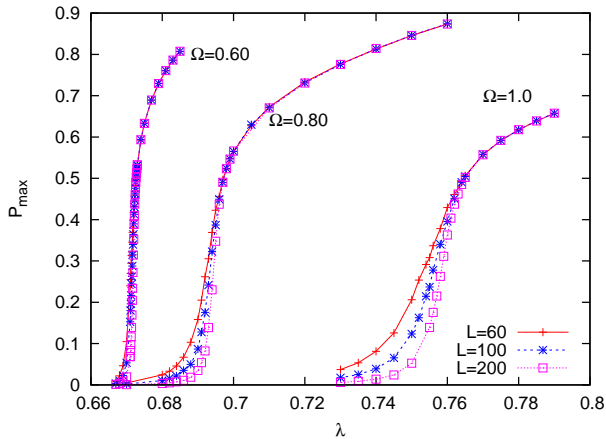}
\caption{(Color online) Comparative plots for the largest cluster size with 
conviction parameter for three 
different system sizes and at three various values of the opinion threshold 
($\Omega = 1.0, 0.80$ and $0.60$).}
\label{comp}

\end{figure}

\begin{figure}

\noindent \includegraphics[clip,width= 6cm, angle=270]{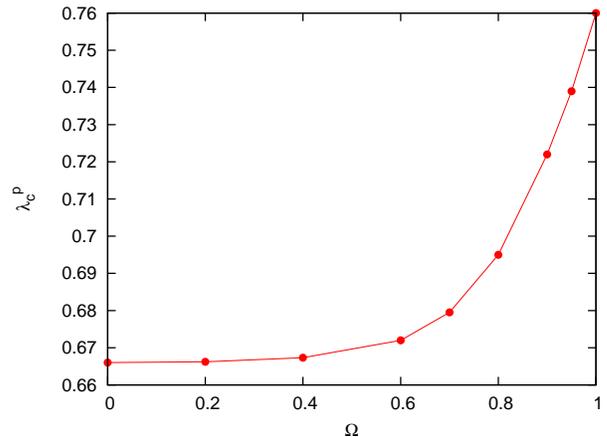}
\caption{(Color online) Plot of the critical conviction parameter ($\lambda_c^p$) 
with the threshold opinion value ($\Omega$).}
\label{crit}

\end{figure}

In this section we will discuss about the percolation behaviour of the geometrical
clusters formed on a square lattice LCCC model. Here we assume that the agents are
placed on the sites of a square lattice and follow the LCCC dynamics. We
define a geometrical cluster as consisting of the adjacent sites having opinion value 
more than or equal to a prefixed threshold opinion 
value ($\Omega$). In our numerical simulation we have used random sequential 
updating rule.
For each value of $\lambda$ and $\Omega$, when the system reaches a 
steady state 
we measure the percolation order parameter $P_{max} = S_L/L^2$ (where $S_L$ is
the size of the largest cluster and $L$ is the linear size of the
system). The value of $P_{max}$ increases with $\lambda$ and at some $\lambda_c^p$ the
system undergoes a percolation transition (Fig.~\ref{fig:fg1}). The value
of $\lambda_c^p$ decreases with decrease in $\Omega$, approaching the value $\lambda_c$
as $\Omega \rightarrow 0.0$ (Figs.~\ref{comp} and ~\ref{crit}). Moreover it is also evident from 
Fig.~\ref{comp} that the finite size effect diminishes with decrease in $\Omega$. 

The percolation transition is characterised by power-law variation of 
different quantities. The order parameter which means the relative size ($P_{max}$) 
of the largest cluster varies as 
\begin{equation}
P_{max}\sim (\lambda_c^p-\lambda)^{\beta}.
\end{equation}

\noindent and the correlation length diverges near the percolation transition point as
\begin{equation}
\xi \sim (\lambda_c^p-\lambda)^{-\nu},
\end{equation}
\noindent where, $\lambda_c^p$ is the critical conviction parameter.
The values of the critical exponents $\beta$ and $\nu$  specify the 
universality class of the transition.

\begin{figure}

\noindent \includegraphics[clip,width= 8cm, angle=0]{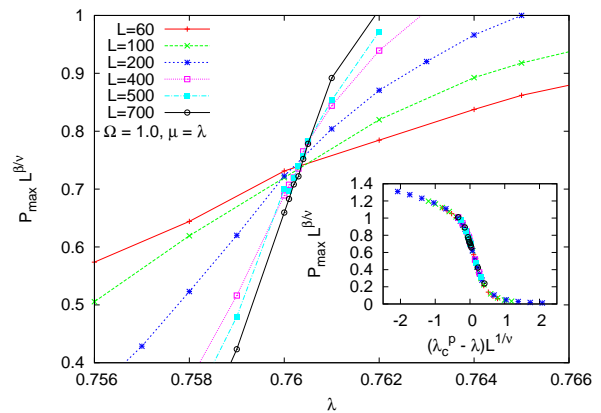}
\caption{(Color online) $P_{max}L^{\beta/\nu}$ plotted against the conviction 
parameter $\lambda$
for $\Omega = 1.0$ and $\mu = \lambda$. The curves for different system sizes
($L = 60, 100, 200, 400, 500$ and $700$) cross at $\lambda_c^p = 0.760\pm0.001$ for 
$\beta/\nu = 0.130\pm0.005$. In the inset the data collapse for $P_{max}$ with 
($\lambda_c^p-\lambda$) has been shown for $\Omega = 1.0$
giving $1/\nu = 0.80\pm0.01$ and $\beta/\nu = 0.130\pm0.005$.}
\label{exc2dperc.th1.0.crossing}

\end{figure}

However, the exponents are not determined from these definitions due to finite size
effects. The critical exponents are determined from the finite size scaling 
relations \cite{Tsakiris,Biswas}. For example, the order parameter is expected to 
follow the scaling form
\begin{equation}
P_{max}=L^{-\beta/\nu}\mathcal{F}\left[L^{1/\nu}\left (\lambda_c^p-\lambda\right)\right],
\end{equation}
\noindent where $\mathcal{F}$ is a suitable scaling function. If we plot
$P_{max}L^{\beta/\nu}$ against $\lambda$ for different system sizes but fixed $\Omega$, then by tuning
$\beta/\nu$, all the curves can be made to cross at a single point. The value of
$\lambda$ for which this happens is the critical conviction parameter 
($\lambda_c^p$). To
estimate $\nu$, $P_{max}L^{\beta/\nu}$ is to be plotted against
$(\lambda_c^p-\lambda)L^{1/\nu}$ and by tuning $1/\nu$, the curves are made to
collapse, giving an accurate estimate of the exponent $\nu$. The other exponents can 
be obtained from scaling relations \cite{Stauffer}.

\begin{figure}

\noindent \includegraphics[clip,width= 8cm, angle=0]{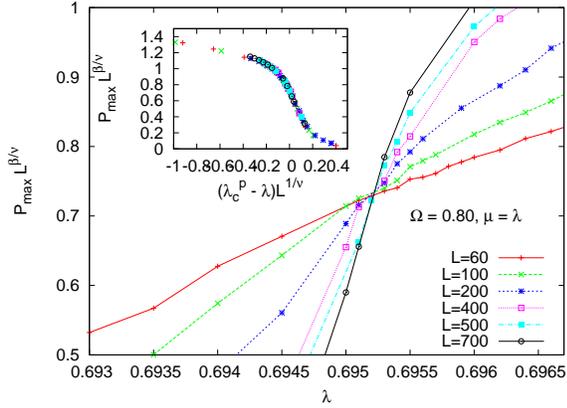}
\caption{(Color online) $P_{max}L^{\beta/\nu}$ plotted against the conviction 
parameter $\lambda$
where $\Omega = 0.80$ and $\mu = \lambda$. The curves for different system sizes
($L = 60, 100, 200, 400, 500$ and $700$) cross at $\lambda_c^p = 0.6955\pm0.0005$ for
$\beta/\nu = 0.130\pm0.005$. In the inset the data collapse for $P_{max}$ with 
$\lambda_c^p-\lambda$ has been shown for $\Omega = 0.80$
giving $1/\nu = 0.80\pm0.01$ and $\beta/\nu = 0.130\pm0.005$.}
\label{exc2dperc.th0.80.crossing}

\end{figure}

\begin{figure}

\noindent \includegraphics[clip,width= 6cm, angle=270]{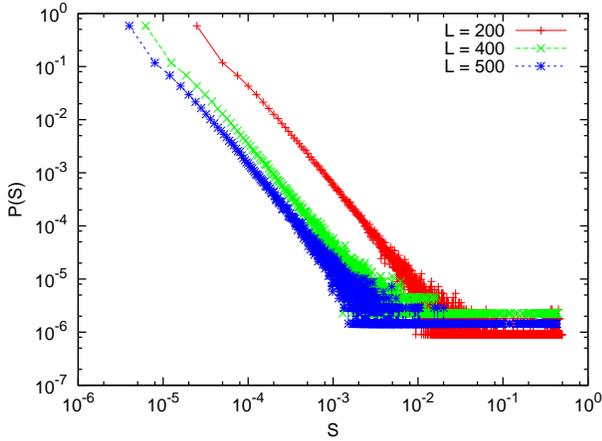}
\caption{(Color online) The cluster size distribution for LCCC model for three 
different system 
sizes at $\Omega = 1.0$ and corresponding critical conviction parameter 
$\lambda_c^p = 0.760$. All the curves decay algebraically 
with an exponent $1.82\pm0.01$.}
\label{clsize}

\end{figure}

For $\Omega=1.0$, we plot $P_{max}L^{\beta/\nu}$ against $\lambda$ 
(Fig.~\ref{exc2dperc.th1.0.crossing}). The curves for different system sizes
($L = 60, 100, 200, 400, 500$ and $700$)
cross at a point when $\beta/\nu = 0.130\pm0.005$ and the crossing point
($\lambda_c^p=0.760\pm0.001$) gives the critical conviction parameter.
Now to determine $\nu$, we plot $P_{max}L^{\beta/\nu}$ against 
$(\lambda_c^p-\lambda)L^{1/\nu}$ and by tuning the value of $1/\nu$ all the three
plots are made to collapse on a single curve (inset of 
Fig.~\ref{exc2dperc.th1.0.crossing}) giving
an estimate of $1/\nu = 0.80\pm0.01$. 
Although with decrease of $\Omega$, the critical point for percolation approaches
$\lambda_c$, the exponents remain same. We have shown the same plots for
$\Omega = 0.80$ in Fig.~\ref{exc2dperc.th0.80.crossing}. The corresponding critical 
point
is $\lambda_c^p=0.6955\pm0.0005$, but $\beta/\nu = 0.130\pm0.005$ and 
$1/\nu = 0.80\pm0.01$. The values of $\beta/\nu$ and $1/\nu$ are different from
that obtained for the percolation transition in case of static Ising 
($\beta_s/\nu_s = 0.052\pm0.002$, $1/\nu_s = 0.996\pm0.009$) \cite{Santo1}, 
dynamic Ising 
($\beta_d/\nu_d = 0.20\pm0.05$, $1/\nu_d = 0.85\pm0.05$) \cite{Biswas} and standard 
percolation 
($\beta/\nu = 5/48,1/\nu = 3/4$) \cite{Stauffer} for two dimensional system.

\begin{figure}

\noindent \includegraphics[clip,width= 8cm, angle=0]{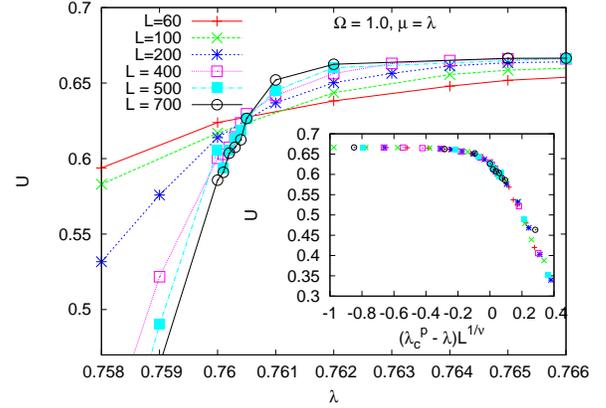}
\caption{(Color online) Fourth-order reduced Binder cumulant of percolation order 
parameter
($P_{max}$) for six different system sizes ($L = 60, 100, 200, 400, 500$ and $700$) 
at $\Omega = 1.0$ and $\mu = \lambda$;
the crossing point determines the critical point ($\lambda_c^p = 0.760\pm0.001$). The
critical Binder cumulant value is $U^{\star}=0.62\pm0.01$. Inset shows the data
collapse for the same value of $1/\nu$ as obtained for the data collapse of $P_{max}$.}
\label{exc2dperc.th1.0.binder}

\end{figure}

\begin{figure}

\noindent \includegraphics[clip,width= 8cm, angle=0]{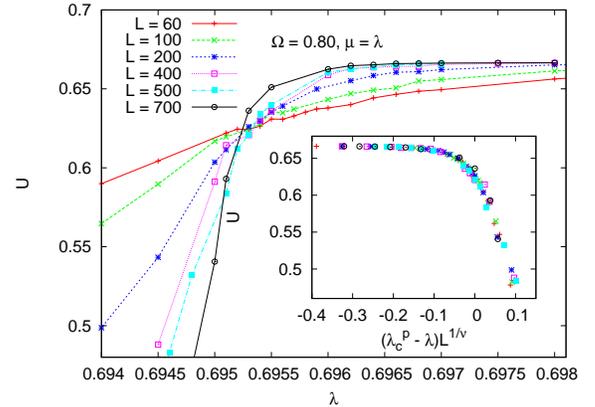}
\caption{(Color online) Fourth-order reduced Binder cumulant of percolation order 
parameter
($P_{max}$) for different system sizes ($L = 60, 100, 200, 400, 500$ and $700$) at 
$\Omega = 0.80$ and $\mu = \lambda$;
the crossing point determines the critical point ($\lambda_c^p = 0.6955\pm0.0005$). 
The critical Binder cumulant value is $U^{\star}=0.62\pm0.01$. Inset shows the data
collapse for the same value of $1/\nu$ as obtained for the data collapse of $P_{max}$.}
\label{exc2dperc.th0.80.binder}

\end{figure}

We have also studied the cluster size distribution for a fixed value of $\Omega$ and
for three different system sizes ($L = 200, 400$ and $500$). For $\Omega = 1.0$, at 
the critical point (i.e. $\lambda = 0.760$) all the curves decay algebraically as 
$P(S) \sim S^{-\tau}$ (where $S$ denotes the sizes of the cluster) with 
$\tau = 1.82\pm0.01$ (Fig.~\ref{clsize}). The value of $\tau$ 
remains same for other values of $\Omega$ (at corresponding critical values of 
$\lambda$). 

For further verification of the critical point and the universality class, we have 
studied
the reduced fourth-order Binder cumulant of the order parameter, defined as 
\cite{binder2}

\begin{equation}
U=1-\frac{\langle P_{max}^4\rangle}{3\langle P_{max}^2\rangle^2},
\end{equation}   

\noindent where $P_{max}$ is the percolation order parameter (as defined before) and 
the angular brackets denote ensemble average. $U \to \frac{2}{3}$ deep inside the 
ordered phase and $U\to 0$ in the disordered phase when the fluctuation is Gaussian.
The crossing point of the different curves ($U-\lambda$) for different system sizes
gives the critical point ($\lambda_c^p=0.760\pm0.001$) for $\Omega=1.0$, which is in 
good 
agreement with the previous estimation from finite size scaling of the corresponding
$\Omega$ (Fig.~\ref{exc2dperc.th1.0.binder}). The value of $U$ at the critical point 
for any value of $\Omega$ is 
$U^{\star}= 0.624\pm0.002$ (see Fig.~\ref{exc2dperc.th1.0.binder}).
The Binder cumulant also follows the scaling form
\begin{equation}
\label{binder-nu}
U=\mathcal{U}((\lambda_c^p-\lambda)L^{1/\nu}),
\end{equation}
\noindent where $\mathcal{U}$ is a suitable scaling function. The data collapse
for $\Omega=1.0$ has been shown in the inset of Fig.~\ref{exc2dperc.th1.0.binder}. 
and the value of $1/\nu$ is $0.80\pm0.01$ which is in good 
agreement with the value of $1/\nu$ obtained from the finite size scaling of the 
largest cluster size. The same plot has been shown in Fig.~\ref{exc2dperc.th0.80.binder}
for $\Omega = 0.80$, which also gives the same value of $1/\nu$, which indicates 
that the critical exponents are independent of $\Omega$.
\linebreak

\begin{figure}

\noindent \includegraphics[clip,width= 8cm, angle=0]{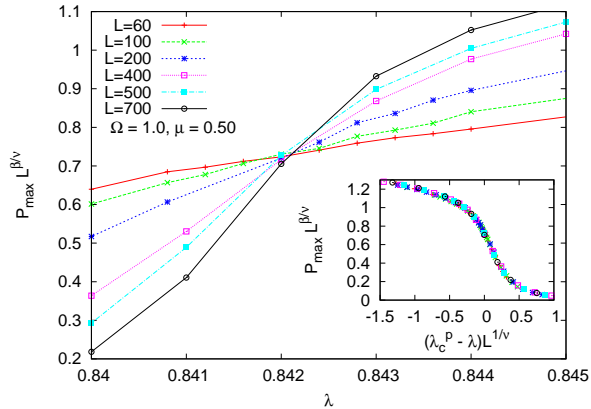}
\caption{(Color online) $P_{max}L^{\beta/\nu}$ plotted against the conviction 
parameter $\lambda$
where $\Omega = 1.0$ and $\mu = 0.50$. The curves for different system sizes
($L = 60, 100, 200, 400, 500$ and $700$) cross at $\lambda_c^p = 0.842\pm0.001$ for
$\beta/\nu = 0.130\pm0.005$. In the inset we have shown the data collapse for 
$P_{max}$ with $\lambda_c^p-\lambda$ for $\Omega = 1.0$
and $\mu = 0.50$ giving $1/\nu = 0.80\pm0.01$ and $\beta/\nu = 0.130\pm0.005$.}
\label{exc2dperc3.th1.0.mu0.50.clus}

\end{figure}

\begin{figure}

\noindent \includegraphics[clip,width= 8cm, angle=0]{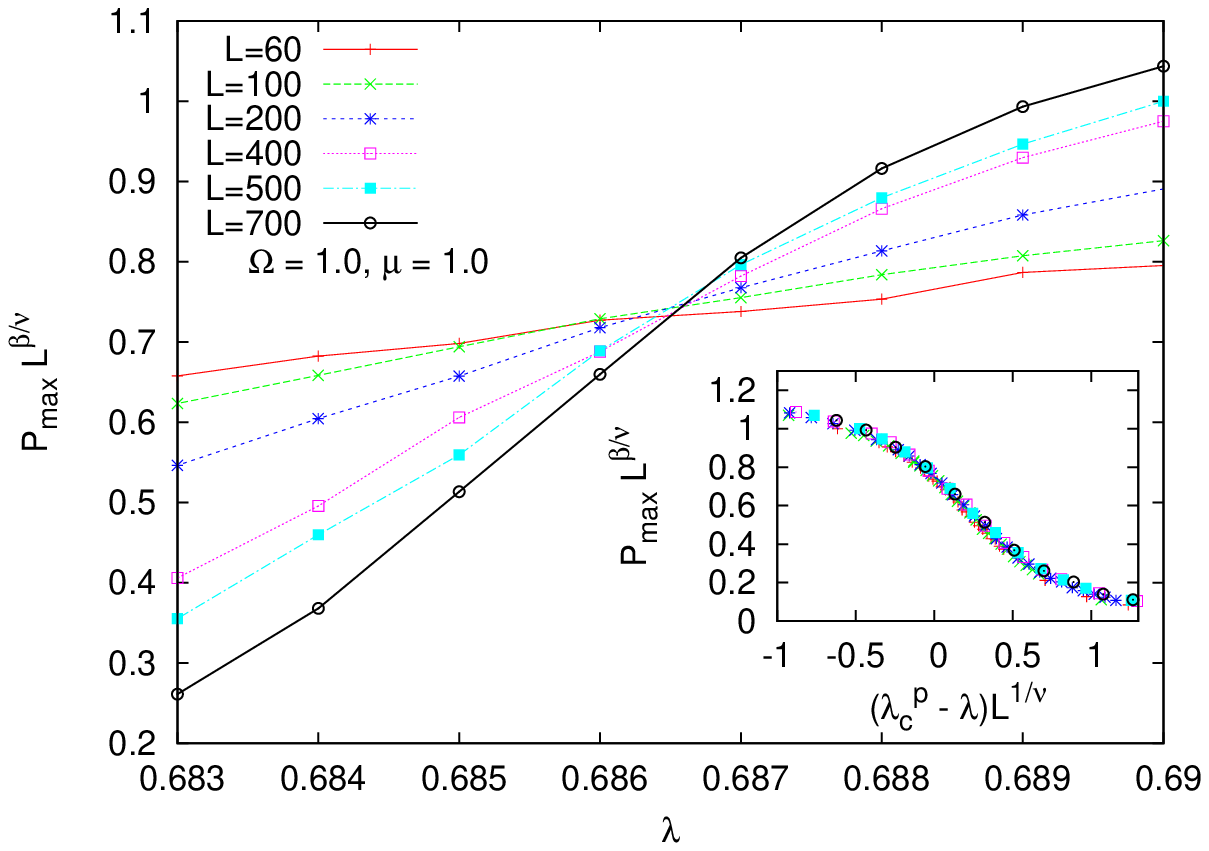}
\caption{(Color online) $P_{max}L^{\beta/\nu}$ plotted against the conviction 
parameter $\lambda$ where $\Omega = 1.0$ and $\mu = 1.0$. The curves for different 
system sizes
($L = 60, 100, 200, 400, 500$ and $700$) cross at $\lambda_c^p = 0.687\pm0.001$ for
$\beta/\nu = 0.130\pm0.005$. In the inset we have shown the data collapse for 
$P_{max}$ with 
$\lambda_c^p-\lambda$ for $\Omega = 1.0$
and $\mu = 1.0$ giving $1/\nu = 0.80\pm0.01$ and $\beta/\nu = 0.130\pm0.005$.}
\label{exc2dperc3.th1.0.mu1.00.clus}

\end{figure}

\begin{figure}

\noindent \includegraphics[clip,width= 8cm, angle=0]{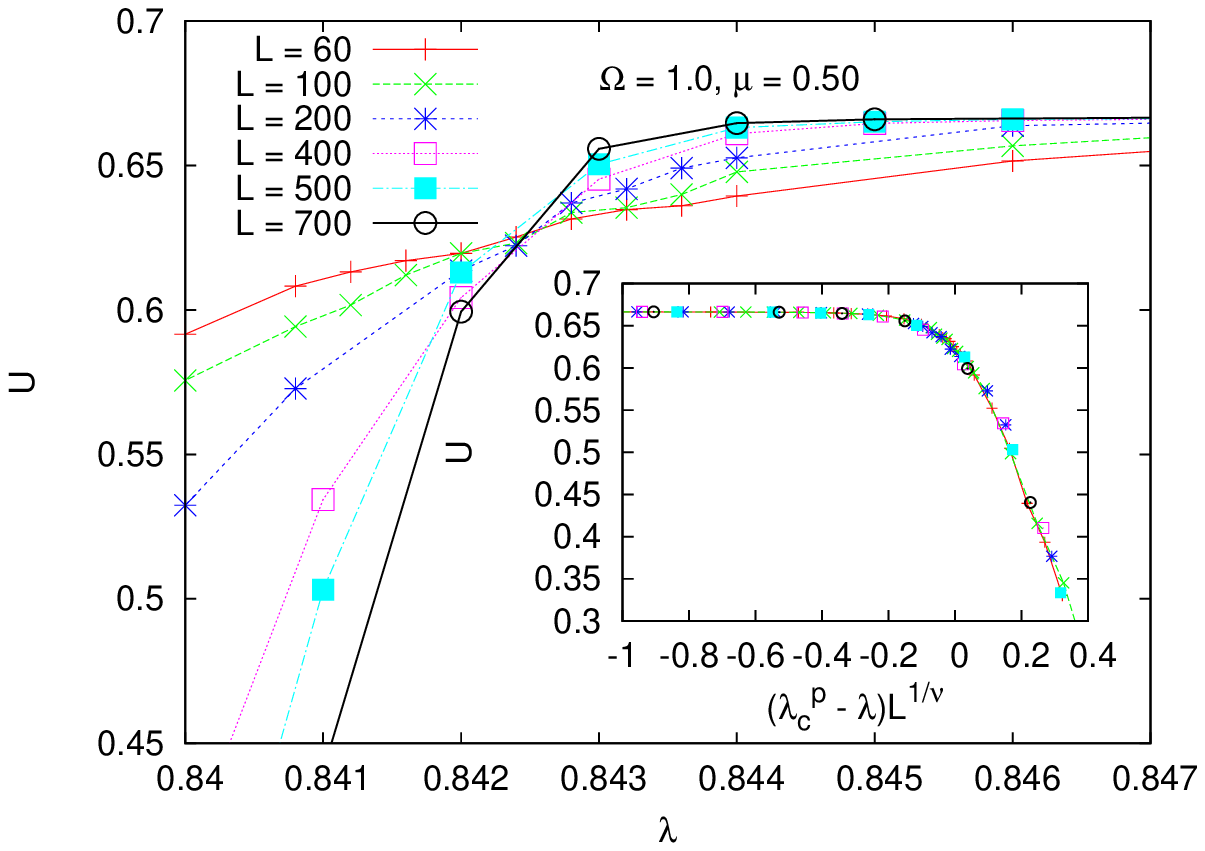}
\caption{(Color online) Fourth-order reduced Binder cumulant of percolation order 
parameter
($P_{max}$) for different system sizes ($L = 60, 100, 200, 400, 500$ and $700$) at 
$\Omega = 1.0$ and $\mu = 0.50$;
the crossing point determines the critical point ($\lambda_c^p = 0.842\pm0.001$) The
critical Binder cumulant value is $U^{\star}=0.624\pm0.002$. Inset shows the data
collapse for the same value of $1/\nu$ as obtained for the data collapse of $P_{max}$.}
\label{exc2dperc3.th1.0.mu0.50.binder}

\end{figure}

\begin{figure}

\noindent \includegraphics[clip,width= 8cm, angle=0]{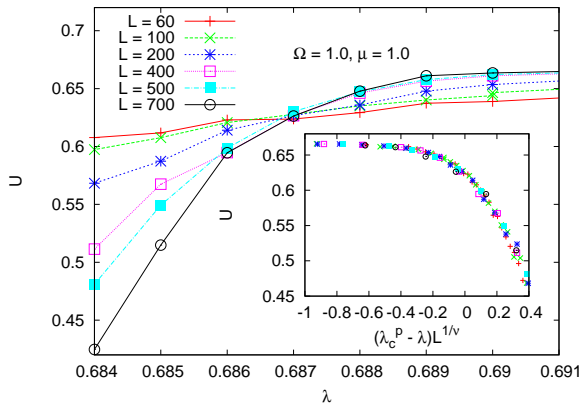}
\caption{(Color online) Fourth-order reduced Binder cumulant of percolation order 
parameter
($P_{max}$) for different system sizes ($L = 60, 100, 200, 400, 500$ and $700$) at 
$\Omega = 1.0$ and $\mu = 1.0$;
the crossing point determines the critical point ($\lambda_c^p = 0.687\pm0.001$) The
critical Binder cumulant value is $U^{\star}=0.624\pm0.002$. Inset shows the data
collapse for the same value of $1/\nu$ as obtained for the data collapse of $P_{max}$.}
\label{exc2dperc3.th1.0.mu1.00.binder}

\end{figure}

%%%%%%%%%%%%%%%%%%%%%%%%%%%%%%%%%%%%%%%%%%%%%%%%%%%%%%%%%%%%%%%%%%%%%%%
\section{Percolation in generalised LCCC model}
%%%%%%%%%%%%%%%%%%%%%%%%%%%%%%%%%%%%%%%%%%%%%%%%%%%%%%%%%%%%%%%%%%%%%%%

We have also investigated the percolation transition in case of generalised LCCC model
in which the conviction parameter ($\lambda$) and the influencing parameter ($\mu$)
are different. We have studied the percolation transition for two sets of parameters :
$\Omega = 1.0,\mu = 0.50$ and $\Omega = 1.0,\mu = 1.0$. In both of the cases the 
plots of $P_{max}L^{\beta/\nu}$ with $\lambda$
for different system sizes ($L = 60, 100, 200, 400, 500$ and $700$) cross at 
a single point for
$\beta/\nu = 0.130\pm0.005$ (Figs.~\ref{exc2dperc3.th1.0.mu0.50.clus} and 
\ref{exc2dperc3.th1.0.mu1.00.clus}) which is same
as obtained for the LCCC model. The critical points are different
($\lambda_c^p = 0.842\pm0.001$ for $\mu = 0.50$ and $\lambda_c^p = 0.687\pm0.001$ for 
$\mu = 1.0$). The value of $1/\nu$ is obtained from the finite size
scaling of the largest cluster size (inset of Fig.~\ref{exc2dperc3.th1.0.mu0.50.clus} 
and \ref{exc2dperc3.th1.0.mu1.00.clus}) and
the estimated values of $\beta/\nu = 0.130\pm0.005$ and $1/\nu = 0.80\pm0.01$ are the same as that obtained for the
LCCC model and are independent of the value of $\mu$, which is in contrary with the
results obtained for the transition of the average opinion value, where the critical
exponents change with $\mu$. This implies that 
the percolation transition is much more robust than the average opinion transition. 
The plots for the Binder cumulant also satisfy the crossing point
and the critical exponents as obtained previously
(Figs.~\ref{exc2dperc3.th1.0.mu0.50.binder} and \ref{exc2dperc3.th1.0.mu1.00.binder}).
The cluster size distribution also decays algebraically with an exponent 
$1.82\pm0.01$ for $\mu = 0.50$ which is the same as that obtained for the LCCC model.

%%%%%%%%%%%%%%%%%%%%%%%%%%%%%%%%%%%%%%%%%%%%%%%%%%%%%%%%%%%%%%%%%%%%%%%
\section{Discussion}
%%%%%%%%%%%%%%%%%%%%%%%%%%%%%%%%%%%%%%%%%%%%%%%%%%%%%%%%%%%%%%%%%%%%%%%

We have investigated the geometrical percolation transition of square lattice LCCC 
model and have found the critical points and the critical exponents 
($\beta/\nu = 0.130\pm0.005, 1/\nu=0.80\pm0.01, \tau=1.82\pm0.01$) characterising the 
transition. Although the system does not show any finite 
size effect in case of the
transition of the average opinion, the percolation transition shows
prominent finite size effect for a given threshold opinion value ($\Omega$).
The finite-size effect diminishes gradually as we decrease the value of $\Omega$. 
The transition point also decreases with $\Omega$ but the change is continuous.
The critical exponents are independent
of the value of the threshold opinion value as well as the value of the conviction
and influencing parameter which shows the robustness of this percolation transition 
in this system. The critical exponents are significantly different from those 
obtained 
in case of static and dynamic Ising system and standard percolation. These 
exponents suggest that this LCCC model belongs to a separate universality class
from the viewpoint of percolation transition. 

\begin{acknowledgements}
The author acknowledges many fruitful discussions and suggestions of 
Prof. Bikas K. Chakrabarti, Mr. Soumyajyoti Biswas and Dr. Arnab Chatterjee. The 
author acknowledges the financial support from
DST (India) under the SERC Fast Track Scheme for Young 
Scientists Sanc. No. SR/FTP/PS-090/2010(G). 
The computational facilities of CAMCS of SINP were also used in producing the 
numerical results.
\end{acknowledgements}


\begin{references}

\bibitem{Stauffer}D. Stauffer and A. Aharony, {\em Introduction to Percolation Theory} (Taylor \& Francis, London, 1994).

\bibitem{Grimmett}G. Grimmett, {\em Percolation} (Springer-Verlag, Berlin, 1999).

\bibitem{Muller} H. M{\"u}ller-Krumbhaar, Phys. Lett. A {\bf 48}, 459 (1974)

\bibitem{Stoll} E. Stoll, K. Binder and T. Schneider, Phys. Rev. B {\bf 6}, 2777 (1972)

\bibitem{Muller2} H. M{\"u}ller-Krumbhaar and E. P. Stoll, J. Chem. Phys. {\bf 65}, 4294 (1976)

\bibitem{Binder} K. Binder and D. Stauffer, J. Stat. Phys. {\bf 6}, 49 (1972)

\bibitem{Coniglio1} A. Coniglio, C. R. Nappi, F. Peruggi and L. Russo, Commun. Math. 
Phys. {\bf 51}, 315 (1976)

\bibitem{Coniglio2} A. Coniglio and W. Klein, J. Phys. A {\bf 13}, 2775 (1980)

\bibitem{Heermann} D. W. Heermann and D. Stauffer, Z. Physik B {\bf 44}, 339 (1981)

\bibitem{Santo1}S. Fortunato, Phys. Rev. B {\bf 66}, 054107 (2002)

\bibitem{Santo2}S. Fortunato, Phys. Rev. B {\bf 67}, 014102 (2003)

\bibitem{Biswas} S. Biswas, A. Kundu and A. K. Chandra, Phys. Rev. E {\bf 83}, 021109 (2011)

\bibitem{Castellano}
C. Castellano, S. Fortunato and V. Loreto, Rev. Mod. Phys. {\bf 81} 591 (2009).

\bibitem{Wiley}{\em Econophysics and Sociophysics: Trends and Perspectives} edited by
B. K. Chakrabarti, A. Chakraborti and A. Chatterjee (Wiley-VCH, Berlin, 2006) 

\bibitem{Holley}R. A. Holley and T. M. Liggett, Ann. Probab. {\bf 3} 643 (1975)

\bibitem{Liggett}T. M. Liggett, {\em Stochastic Interacting Systems: Contact, Voter
and Exclusion Processes} (Springer, Berlin, 1999); R. Lambiotte and S. Redner,
Europhys. Lett. {\bf 82}, 18 007 (2008) 

\bibitem{Sznajd} K. Sznajd-Weron and J. Sznajd, Int. J. Mod. Phys. C {\bf 11}, 1157
(2000)

\bibitem{Hegselmann} R. Hegselmann and U. Krause, J. Artif. Soc. Soc. Simul. {\bf 5}, 3(2002)

\bibitem{Deffuant}G. Deffuant, D. Neau, D. Amblard and G. Weisbuch, Adv. Complex Syst. {\bf 3}, 87 (2000)

\bibitem{Fortunato1}S. Fortunato, Int. J. Mod. Phys. C {\bf 16}, 17 (2005)

\bibitem{BCS}S. Biswas, A. Chatterjee and P. Sen, arXiv:1102.0902v3 ; to appear in
Physica A. 

\bibitem{Lallouache1}
M. Lallouache, A. Chakraborti and B. K. Chakrabarti, Sci. Cult. {\bf 76} (9-10), 485 
(2010)

\bibitem{Anirban} A. Chakraborti and B. K. Chakrabarti, in {\em Econophysics of
order-driven markets} edited by F. Abergel, B. K. Chakrabarti, A. Chakraborti
and M. Mitra (Springer-Verlag, Milan, 2011)

\bibitem{Shao}J. Shao, S. Havlin and H. E. Stanley, Phys. Rev. Lett. {\bf 103}, 018701
(2009)

\bibitem{Lallouache2}
M. Lallouache, A. S. Chakrabarti, A. Chakraborti and B. K. Chakrabarti, Phys. Rev. E
{\bf 82}, 056112 (2010)

\bibitem{Biswas2} S. Biswas, A. K. Chandra, A. Chatterjee and B. K. Chakrabarti,
J. Phys. Conf. Ser. {\bf 297}, 012004 (2011)

\bibitem{Sen} P. Sen, Phys. Rev. E {\bf 83}, 016108 (2011)

\bibitem{Biswas3} S. Biswas, Phys. Rev. E {\bf 84}, 056106 (2011)

\bibitem{Chakrabarti}
A. Chakraborti and B. K. Chakrabarti, Eur. Phys. J. B {\bf 17}, 167 (2000); A. 
Chatterjee, B. K. Chakrabarti and S. S. Manna Physica A {\bf 335}, 155 (2004); A. 
Chatterjee and B. K. Chakrabarti, Eur. Phys. J. B {\bf 60}, 135 (2007)

\bibitem{Tsakiris} N. Tsakiris, M. Maragakis, K. Kosmidis, and P. Argyrakis, Phys. Rev. E {\bf 82}, 041108 (2010)

\bibitem{binder2}
K. Binder and D. Heermann, {\em Monte Carlo Simulations in Statistical Physics} 
(Springer, Berlin, 1988).



\end{references}
\end{document}